\documentclass[12pt]{article}
\usepackage[a4paper]{geometry}
\usepackage[myheadings]{fullpage}
\usepackage{fancyhdr}
\usepackage{lastpage}
\usepackage{graphicx, wrapfig, subcaption, setspace, booktabs}
\usepackage[T1]{fontenc}
\usepackage[font=small, labelfont=bf]{caption}
\usepackage{fourier}
\usepackage[protrusion=true, expansion=true]{microtype}
\usepackage[english]{babel}
\usepackage{sectsty}
\usepackage{todonotes}
\usepackage{url, lipsum}

\newcommand{\HRule}[1]{\rule{\linewidth}{#1}}
\begin{document}

\title{ \LARGE \textbf{Content-Centric Networking}
        \HRule{2pt} \\ [0.5cm]
        \large Architectural Overview and Protocol Description}

\date{\today \\ Version: 1.0}

\author{Marc Mosko\\
        PARC \\
        marc.mosko@parc.com
        \and
        Ignacio Solis\footnote{Work done while at PARC.} \\
        LinkedIn \\
        nsolis@linkedin.com
        \and
        Christopher A. Wood \\
        University of California Irvine \\
        woodc1@uci.edu
}

\maketitle
\tableofcontents
\newpage

\section{Overview}
CCNx is a request and response protocol to fetch chunks of data using
a name.  The integrity of each chunk may be directly asserted through
a digital signature or message authentication code (MAC), or,
alternatively, indirectly via hash chains.  Chunks may also carry
weaker message integrity checks (MICs) or no integrity protection
mechanism at all.  Because provenance information is carried with
each chunk (or larger indirectly protected block), we no longer need
to rely on host identities, such as those derived from TLS
certificates, to ascertain the chunk legitimacy.  Data integrity is
therefore a core feature of CCNx; it does not rely on the data
transmission channel.  There are several options for data
confidentiality, discussed later.

As a request and response protocol, CCNx may be carried over many
different transports.  In use today are Ethernet, TCP, UDP, 802.15.4,
GTP, GRE, DTLS, TLS, and others.  While the specific wire format of
CCNx may vary to some extent based on transport, the core principles
and behaviors of CCNx outlined in this document should remain fixed.

CCNx uses hierarchical names to identify bytes of payload.  The Name
combines a routable prefix with an arbitrary application-dependent
suffix assigned by the publisher to a piece of content.  The result
is a "named payload".  This is different from other systems that use
only self-certifying names, where the payload name is intrinsically
derivable from the payload or its realization in a network object
(e.g., a SHA-256 hash of the payload or network object).  In human-
readable form, we represent names as a "ccnx:" scheme URI
\cite{berners2014rfc}, though the canonical encoding should be octet strings.  In
this respect, we speak of a name being made up of hierarchical path
segments, which is the URI terminology.

This document only defines the general properties of CCNx names.  In
some isolated environments, CCNx users may be able to use any name
they choose and either inject that name (or prefix) into a routing
protocol or use other information foraging techniques. In the
Internet environment, there will be policies around the formats of
names and assignments of names to publishers, though those are not
specified here.

The key concept of CCNx is that a subjective name is
(cryptographically) bound to a fixed payload. These (publisher-
generated) bindings can therefore be (cryptographically) verified.
For example, a publisher could compute a cryptographic hash over the
name and payload, sign the hash, and deliver the tuple {Name,
Payload, Validation}.  Consumers of this data can check the binding
integrity by re-computing the same cryptographic hash and verifying
the digital signature in Validation.  Additional information would be
included as needed by specific validation mechanisms.  Therefore, we
divide Validation in to a ValidationAlgorithm and a
ValidationPayload.  The ValidationAlgorithm has information about the
crypto suite and parameters.  In particular, the ValidationAlgorithm
usually has a field called KeyId which identifies the public key used
by the validation, when applicable.  The ValidationPayload is the
output of the validation algorithm, such as a CRC value, an HMAC
output, or an RSA signature.

In addition to the essential Name, Payload, and Validation sections,
a CCNx user may need to include some other signaling information.
This could include a hint about the type of Payload (e.g.,
application data, a cryptographic key, etc.) or cache control
directives, etc.  We will call this extra signaling information
ExtraFields.

A named payload is thus the nested tuple
$$
((\mathsf{Name}, \mathsf{ExtraFields}, \mathsf{Payload}, \mathsf{ValidationAlgorithm}), \mathsf{ValidationPayload}),
$$
where all fields in the inner tuple are covered by the value in the validation payload.

CCNx specifies a network protocol around Interests (request messages)
and Content Objects (response messages) to move named payloads. An
Interest includes the Name -- which identifies the desired response
-- and two optional limiting restrictions.  The first restriction on
the KeyId to limit responses to those signed with a
ValidationAlgorithm KeyId field equal to the restriction. The second
is the ContentObjectHash restriction, which limits the response to
one where the cryptographic hash of the entire named payload is equal
to the restriction.

The hierarchy of a CCNx Name is used for routing via the longest
matching prefix in a Forwarder. The longest matching prefix is
computed name segment by name segment in the hierarchical path name,
where each name segment must be exactly equal to match.  There is no
requirement that the prefix be globally routable.  Within a
deployment any local routing may be used, even one that only uses a
single flat (non-hierarchical) name segment.

Another concept of CCNx is that there should be flow balance between
Interest messages and Content Object messages. At the network level,
an Interest traveling along a single path should elicit no more than
one Content Object response. If some node sends the Interest along
more than one path, that node should consolidate the responses such
that only one Content Object flows back towards the requester.  If an
Interest is sent broadcast or multicast on a multiple-access media,
the sender should be prepared for multiple responses unless some
other media-dependent mechanism like gossip suppression or leader
election is used.

As an Interest travels the forward path following the Forwarding
Information Base (FIB), it establishes state at each forwarder such
that a Content Object response can trace its way back to the original
requester(s) without the requester needing to include a routable
return address.  We use the notional Pending Interest Table (PIT) as
a method to store state that facilitates the return of a Content
Object. The PIT table is not mandated by the specification.

The notional PIT table stores the last hop of an Interest plus its
Name and optional restrictions.  This is the data required to match a
Content Object to an Interest (see Section \ref{sec:matching}).  When a Content Object
arrives, it must be matched against the PIT to determine which
entries it satisfies.  For each such entry, at most one copy of the
Content Object is sent to each listed last hop in the PIT entries.

If multiple Interests with the same {Name, KeyIdRestriction,
ContentObjectHashRestriction} tuple arrive at a node before a Content
Object matching the first Interest comes back, they are grouped in
the same PIT entry and their last hops aggregated (see
Section 2.4.2).  Thus, one Content Object might satisfy multiple
pending Interests in a PIT.

In CCNx, higher-layer protocols often become so-called "name-based
protocols" because they operate on the CCNx Name.  For example, a
versioning protocol might append additional name segments to convey
state about the version of payload.  A content discovery protocol
might append certain protocol-specific name segments to a prefix to
discover content under that prefix.  Many such protocols may exist
and apply their own rules to Names.  They may be layered with each
protocol encapsulating (to the left) a higher layer's Name prefix.

This document also describes a control message called an
InterestReturn.  A network element may return an Interest message to
a previous hop if there is an error processing the Interest.  The
returned Interest may be further processed at the previous hop or
returned towards the Interest origin.  When a node returns an
Interest it indicates that the previous hop should not expect a
response from that node for the Interest, i.e., there is no PIT entry
left at the returning node for a Content Object to follow.

There are multiple ways to describe larger objects in CCNx.  Some
options may use the namespace while others may use a structure such
as a Manifest.  This document does not address these options at this
time.

The remainder of this document describes a named payload as well as
the Interest and Content Object network protocol behavior in detail.

\section{Protocol}
CCNx is a request and response protocol.  A request is called an
Interest and a response is called a ContentObject.  CCNx also uses a
1-hop control message called InterestReturn.  These are, as a group,
called CCNx Messages.

\subsection{Message Grammar}
The CCNx message ABNF \cite{overell2008augmented} grammar is show in Figure 1.  The
grammar does not include any encoding delimiters, such as TLVs.
Specific wire encodings are given in a separate document.  If a
Validation section exists, the Validation Algorithm covers from the
Body (BodyName or BodyOptName) through the end of the ValidationAlg
section.  The InterestLifetime, CacheTime, and Return Code fields
exist outside of the validation envelope and may be modified.

The various fields -- in alphabetical order -- are defined as:

\begin{itemize}
\item AbsTime: Absolute times are conveyed as the 64-bit UTC time in
      milliseconds since the epoch (standard POSIX time).
\item CacheTime: The absolute time after which the publisher believes
      there is low value in caching the content object.  This is a
      recommendation to caches (see Section 4).
\item ConObjField: These are optional fields that may appear in a
      Content Object.
\item ConObjHash: The value of the Content Object Hash, which is the
   SHA256-32 over the message from the beginning of the body to the
   end of the message.  Note that this coverage area is different
   from the ValidationAlg.  This value SHOULD NOT be trusted across
   domains (see Section 5).
\item ExpiryTime: An absolute time after which the content object should
   be considered expired (see Section 4).
\item HopLimit: Interest messages may loop if there are loops in the
   forwarding plane.  To eventually terminate loops, each Interest
   carries a HopLimit that is decremented after each hop and no
   longer forwarded when it reaches zero.  See Section 2.4.
\item InterestField: These are optional fields that may appear in an
   Interest message.
\item KeyIdRestr: The KeyId Restriction.  A Content Object must have a
   KeyId with the same value as the restriction.
\item ObjHashRestr: The Content Object Hash Restriction.  A content
   object must hash to the same value as the restriction using the
   same HashType.  The ObjHashRestr MUST use SHA256-32.
\item KeyId: An identifier for the key used in the ValidationAlg.  For
   public key systems, this should be the SHA-256 hash of the public
   key.  For symmetric key systems, it should be an identifer agreed
   upon by the parties.
\item KeyLink: A Link (see Section 6) that names how to retrieve the key
   used to verify the ValidationPayload.  A message SHOULD NOT have
   both a KeyLink and a PublicKey.
\item Lifetime: The approximate time during which a requester is willing
   to wait for a response, usually measured in seconds.  It is not
   strongly related to the network round trip time, though it must
   necessarily be larger.
\item Name: A name is made up of a non-empty first segment followed by
    zero or more additional segments, which may be of 0 length.  Path
    segments are opaque octet strings, and are thus case-sensitive if
    encoding UTF-8.  An Interest MUST have a Name.  A ContentObject
    MAY have a Name (see Section 9).  The segments of a name are said
    to be complete if its segments uniquely identify a single Content
    Object.  A name is exact if its segments are complete.  An
    Interest carrying a full name is one which specifies an exact name
    and the ObjHashRestr of the corresponding Content Object.
\item Payload: The message's data, as defined by PayloadType.
\item PayloadType: The format of the Payload.  If missing, assume
   DataType.  DataType means the payload is opaque application bytes.
   KeyType means the payload is a DER-encoded public key.  LinkType
   means it is one or more Links (see Section 6).
\item PublicKey: Some applications may wish to embed the public key used
   to verify the signature within the message itself.  The PublickKey
   is DER encoded.  A message SHOULD NOT have both a KeyLink and a
   PublicKey.
\item RelTime: A relative time, measured in milli-seconds.
\item ReturnCode: States the reason an Interest message is being
   returned to the previous hop (see Section 10.2).
\item SigTime: The absolute time (UTC milliseconds) when the signature
   was generated.
\item Hash: Hash values carried in a Message carry a HashType to
   identify the algorithm used to generate the hash followed by the
   hash value.  This form is to allow hash agility.  Some fields may
   mandate a specific HashType.
\end{itemize}

\begin{figure}
\begin{verbatim}
Message       := Interest / ContentObject / InterestReturn
Interest      := HopLimit [Lifetime] BodyName [Validation]
ContentObject := [CacheTime / ConObjHash] BodyOptName [Validation]
InterestReturn:= ReturnCode Interest
BodyName      := Name Common
BodyOptName   := [Name] Common
Common        := *Field [Payload]
Validation    := ValidationAlg ValidatonPayload

Name          := FirstSegment *Segment
FirstSegment  := 1* OCTET
Segment       := 0* OCTET

ValidationAlg := RSA-SHA256 HMAC-SHA256 CRC32C
ValidatonPayload := 1* OCTET
RSA-SHA256    := KeyId [PublicKey] [SigTime] [KeyLink]
HMAC-SHA256   := KeyId [SigTime] [KeyLink]
CRC32C        := [SigTime]

AbsTime       := 8 OCTET ; 64-bit UTC msec since epoch
CacheTime     := AbsTime
ConObjField   := ExpiryTime / PayloadType
ConObjHash    := Hash ; The Content Object Hash
ExpiryTime    := AbsTime
Field         := InterestField / ConObjField
Hash          := HashType 1* OCTET
HashType      := SHA256-32 / SHA512-64 / SHA512-32
HopLimit      := OCTET
InterestField := KeyIdRestr / ObjHashRestr
KeyId         := 1* OCTET ; key identifier
KeyIdRestr    := 1* OCTET
KeyLink       := Link
Lifetime      := RelTime
Link          := Name [KeyIdResr] [ObjHashRestr]
ObjHashRestr  := Hash
Payload       := *OCTET
PayloadType   := DataType / KeyType / LinkType
PublicKey     := ; DER-encoded public key
RelTime       := 1* OCTET ; msec
ReturnCode    := ; see Section 10.2
SigTime       := AbsTime
\end{verbatim}
\end{figure}

\subsection{Consumer Behavior}
To request a piece of content for a given
$$
(\mathsf{Name}, [\mathsf{KeyIdRest}], [\mathsf{ObjHashRestr}])
$$
tuple, a consumer creates an Interest message with
those values.  It MAY add a validation section, typically only a
CRC32C.  A consumer MAY put a Payload field in an Interest to send
additional data to the producer beyond what is in the Name.  The Name
is used for routing and may be remembered at each hop in the notional
PIT table to facilitate returning a content object; Storing large
amounts of state in the Name could lead to high memory requirements.
Because the Payload is not considered when forwarding an Interest or
matching a Content Object to an Interest, a consumer SHOULD put an
Interest Payload ID (see Section Section 3.2) as part of the name to
allow a forwarder to match Interests to content objects and avoid
aggregating Interests with different payloads.  Similarly, if a
consumer uses a MAC or a signature, it SHOULD also include a unique
segment as part of the name to prevent the Interest from being
aggregated with other Interests or satisfied by a Content Object that
has no relation to the validation.

The consumer SHOULD specify an InterestLifetime, which is the length
of time the consumer is willing to wait for a response.  The
InterestLifetime is an application-scale time, not a network round
trip time (see Section 2.4.2).  If not present, the InterestLifetime
will use a default value (TO\_INTERESTLIFETIME).

The consumer SHOULD set the Interest HopLimit to a reasonable value
or use the default 255.  If the consumer knows the distances to the
producer via routing, it SHOULD use that value.

A consumer hands off the Interest to its first forwarder, which will
then forward the Interest over the network to a publisher (or
replica) that may satisfy it based on the name (see Section 2.4).

Interest messages are unreliable.  A consumer SHOULD run a transport
protocol that will retry the Interest if it goes unanswered, up to
the InterestLifetime.  No transport protocol is specified in this
document.

The network MAY send to the consumer an InterestReturn message that
indicates the network cannot fulfill the Interest.  The ReturnCode
specifies the reason for the failure, such as no route or congestion.
Depending on the ReturnCode, the consumer MAY retry the Interest or
MAY return an error to the requesting application.

If the content was found and returned by the first forwarder, the
consumer will receive a ContentObject.  The consumer SHOULD:

\begin{itemize}
\item Ensure the content object is properly formatted.

\item Verify that the returned Name matches a pending request.  If the
      request also had KeyIdRestr and ObjHashRest, it should also
      validate those properties.

\item If the content object is signed, it SHOULD cryptographically
      verify the signature.  If it does not have the corresponding key,
      it SHOULD fetch the key, such as from a key resolution service or
      via the KeyLink.

\item If the signature has a SigTime, the consumer MAY use that in
      considering if the signature is valid.  For example, if the
      consumer is asking for dynamically generated content, it should
      expect the SigTime to not be before the time the Interest was
      generated.

\item If the content object is signed, it should assert the
      trustworthiness of the signing key to the namespace.  Such an
      assertion is beyond the scope of this document, though one may use
      traditional PKI methods, a trusted key resolution service, or
      methods like schematized trust \cite{yu2015schematizing}.

\item It MAY cache the content object for future use, up to the
      ExpiryTime if present.

\item A consumer MAY accept a content object off the wire that is
      expired.  It may happen that a packet expires while in flight, and
      there is no requirement that forwarders drop expired packets in
      flight.  The only requirement is that content stores, caches, or
      producers MUST NOT respond with an expired content object.
\end{itemize}

\subsection{Publisher Behavior}
This document does not specify the method by which names populate a
Forwarding Information Base (FIB) table at forwarders (see
Section 2.4).  A publisher is either configured with one or more name
prefixes under which it may create content, or it chooses its name
prefixes and informs the routing layer to advertise those prefixes.

When a publisher receives an Interest, it SHOULD:

\begin{itemize}
\item Verify that the Interest is part of the publishers namespace(s).

\item If the Interest has a Validation section, verify the
    ValidationPayload.  Usually an Interest will only have a CRC32C
    unless the publisher application specifically accommodates other
    validations.  The publisher MAY choose to drop Interests that
    carry a Validation section if the publisher application does not
    expect those signatures as this could be a form of computational
    denial of service.  If the signature requires a key that the
    publisher does not have, it is NOT RECOMMENDED that the publisher
    fetch the key over the network, unless it is part of the
    application's expected behavior.

\item Retrieve or generate the requested content object and return it to
    the Interest's previous hop.  If the requested content cannot be
    returned, the publisher SHOULD reply with an InterestReturn or a
    content object with application payload that says the content is
    not available; this content object should have a short ExpiryTime
    in the future.
\end{itemize}

\subsection{Forwarder Behavior}
\begin{figure}
\center
\includegraphics[scale=0.5]{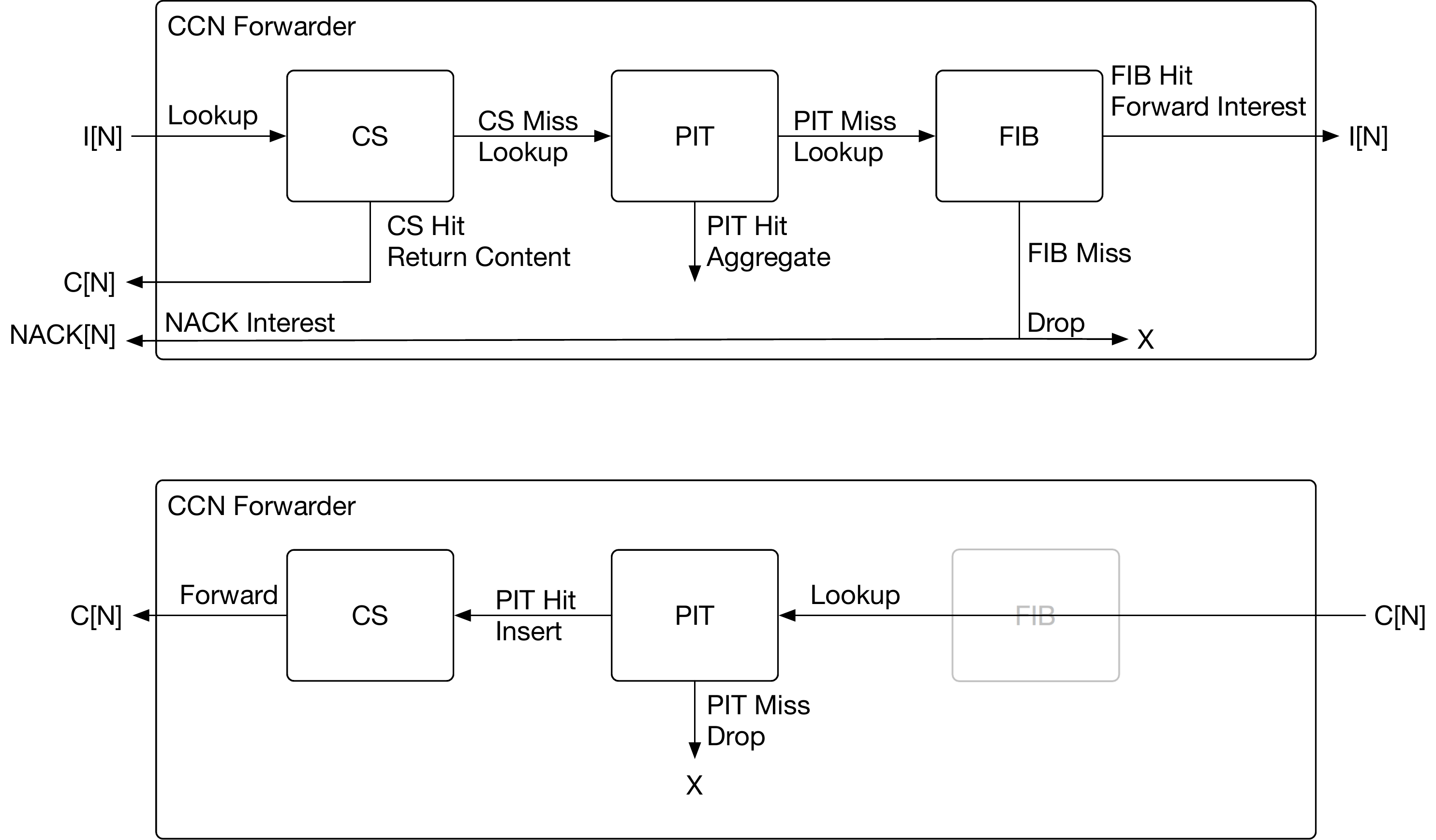}
\caption{The CCN forwarder data path}
\label{fig:datapath}
\end{figure}

A forwarder routes Interest messages based on a Forwarding
Information Base (FIB), returns Content Objects that match Interests
to the Interest's previous hop, and processes InterestReturn control
messages.  It may also keep a cache of Content Objects in the
notional Content Store table.  These functions are shown in Figure \ref{fig:datapath}.
These and other external behaviors are described in the remainder of this section.

In this document, we will use two processing pipelines, one for
Interests and one for Content Objects.  Interest processing is made
up of checking for duplicate Interests in the PIT (see
Section 2.4.2), checking for a cached Content Object in the Content
Store (see Section 2.4.3), and forwarding an Interest via the FIB.
Content Store processing is made up of checking for matching
Interests in the PIT and forwarding to those previous hops.

\subsubsection{Interest HopLimit}
Interest looping is not prevented in CCNx.  An Interest traversing
loops is eventually discarded using the hop-limit field of the
Interest, which is decremented at each hop traversed by the Interest.
Every Interest MUST carry a HopLimit.

When an Interest is received from another forwarder, the HopLimit
MUST be positive.  A forwarder MUST decement the HopLimit of an
Interest by at least 1 before it is forwarded.
If the HopLimit equals 0, the Interest MUST NOT be forwarded to
another forwarder; it MAY be sent to a publisher application or
serviced from a local Content Store.

\subsubsection{Interest Aggregation}
Interest aggregation is when a forwarder receives an Interest message
that could be satisfied by another Interest message already forwarded
by the node so the forwarder suppresses the new Interest; it only
records the additional previous hop so a Content Object sent in
response to the first Interest will satisfy both Interests.

CCNx uses an interest aggregation rule that assumes the
InterestLifetime is akin to a subscription time and is not a network
round trip time.  Some previous aggregation rules assumed the
lifetime was a round trip time, but this leads to problems of
expiring an Interest before a response comes if the RTT is estimated
too short or interfering with an ARQ scheme that wants to re-transmit
an Interest but a prior interest over-estimated the RTT.

A forwarder MAY implement an Interest aggregation scheme.  If it does
not, then it will forward all Interest messages.  This does not imply
that multiple, possibly identical, Content Objects will come back.  A
forwarder MUST still satisfy all pending Interests, so one Content
Object could satisfy multiple similar interests, even if the
forwarded did not suppress duplicate Interest messages.

A RECOMMENDED Interest aggregation scheme is:

\begin{itemize}

\item Two Interests are considered 'similar' if they have the same Name,
   KeyIdRestr, and ObjHashRestr.

\item Let the notional value InterestExpiry (a local value at the
   forwarder) be equal to the receive time plus the InterestLifetime
   (or a platform-dependent default value if not present).

\item An Interest record (PIT entry) is considered invalid if its
   InterestExpiry time is in the past.

\item The first reception of an Interest MUST be forwarded.

\item A second or later reception of an Interest similar to a valid
   pending Interest from the same previous hop MUST be forwarded.  We
   consider these a retransmission requests.

\item A second or later reception of an Interest similar to a valid
   pending Interest from a new previous hop MAY be aggregated (not
   forwarded).

\item Aggregating an Interest MUST extend the InterestExpiry time of the
    Interest record.  An implementation MAY keep a single
    InterestExpiry time for all previous hops or MAY keep the
    InterestExpiry time per previous hop.  In the first case, the
    forwarder might send a ContentObject down a path that is no longer
    waiting for it, in which case the previous hop (next hop of the
    Content Object) would drop it.
\end{itemize}

\subsubsection{Content Store Behavior}
The ContentStore is a special cache that sits on the fast path of a
CCNx forwarder.  It is an optional component.  It serves to repair
lost packets and handle flash requests for popular content.  It could
be pre-populated or use opportunistic caching.  Because the Content
Store could serve to amplify an attach via cache poisoning, there are
special rules about how a Content Store behaves.

\begin{enumerate}
\item  A forwarder MAY implement a ContentStore.  If it does, the
    Content Store matches a Content Object to an Interest via the
    normal matching rules (see Section 9).

\item  If an Interest has a KeyIdRestr, then the ContentStore MUST NOT
    reply unless it knows the signature on the matching ContentObject
    is correct.  It may do this by external knowledge (i.e., in a
    managed system pre-populating the cachine) or by having the
    public key and cryptographically verifying the signature.  If the
    public key is provided in the ContentObject itself (i.e., in the
    PublicKey field) or in the Interest, the ContentStore MUST verify
    that the public key's SHA-256 hash is equal to the KeyId and that
    it verifies the signature.  A ContentStore MAY verify the digital
    signature of a Content Object before it is cached, but it is not
    required to do so.  A ContentStore SHOULD NOT fetch keys over the
    network.  If it cannot or has not yet verified the signature, it
    should treat the Interest as a cache miss.

\item  If an Interest has an ObjHashRestr, then the ContentStore MUST
    NOT reply unless it knows the the matching ContentObject has the
    correct hash.  If it cannot verify the hash, then it should treat
    the Interest as a cache miss.

\item  It must object the Cache Control directives (see Section 4).
\end{enumerate}

\subsubsection{Interest Pipeline}
\begin{enumerate}
\item  Perform the HopLimit check (see Section 2.4.1).

\item Determine if the Interest can be aggregated, as per
     Section 2.4.2.  If it can be, aggregate and do not forward the
     Interest.

\item If forwarding the Interest, check for a hit in the Content Store,
    as per Section 2.4.3.  If a matching Content Object is found,
    return it to the Interest's previous hop.  This injects the
    ContentStore as per Section 2.4.5.

\item Lookup the Interest in the FIB.  Longest prefix match (LPM) is
    performed name segment by name segment (not byte or bit).  It
    SHOULD exclude the Interest's previous hop.  If a match is found,
    forward the Interest.  If no match is found or the forwarder
    choses to not forward due to a local condition (e.g.,
    congestion), it SHOULD send an InterestReturn message, as per
    Section 10.
\end{enumerate}

\subsubsection{Content Object Pipeline}
\begin{enumerate}
\item  It is RECOMMENDED that a forwarder that receives a content object
    check that the ContentObject came from an expected previous hop.
    An expected previous hop is one pointed to by the FIB or one
    recorded in the PIT as having had a matching Interest sent that
    way.

\item  A Content Object MUST be matched to all pending Interests that
    satisfy the matching rules (see Section 9).  Each satisfied
    pending Interest MUST then be removed from the set of pending
    Interests.

\item  A forwarder SHOULD NOT send more then one copy of the received
    Content Object to the same Interest previous hop.  It may happen,
    for example, that two Interest ask for the same Content Object in
    different ways (e.g., by name and by name an KeyId) and that they
    both come from the same previous hop.  It is normal to send the
    same content object multiple times on the same interface, such as
    Ethernet, if it is going to different previous hops.

\item  A Content Object SHOULD only be put in the Content Store if it
    satisfied an Interest (and passed rule \#1 above).  This is to
    reduce the chances of cache poisoning.
\end{enumerate}

\section{Names}
A CCNx name is a composition of name segments.  Each name segment
carries a label identifying the purpose of the name segment, and a
value.  For example, some name segments are general names and some
serve specific purposes, such as carrying version information or the
sequencing of many chunks of a large object into smaller, signed
Content Objects.

There are three different types of names in CCNx: prefix, exact, and
full names.  A prefix name is simply a name that does not uniquely
identify a single Content Object, but rather a namespace or prefix of
an existing Content Object name.  An exact name is one which uniquely
identifies the name of a Content Object.  A full name is one which is
exact and is accompanied by an explicit or implicit ConObjHash.  The
ConObjHash is explicit in an Interest and implicit in a Content
Object.

The name segment labels specified in this document are given in the
table below.  Name Segment is a general name segment, typically
occurring in the routable prefix and user-specified content name.
Other segment types are for functional name components that imply a
specific purpose.

A forwarding table entry may contain name segments of any type.
Routing protocol policy and local system policy may limit what goes
into forwarding entries, but there is no restriction at the core
level.  An Interest routing protocol, for example, may only allow
binary name segments.  A load balancer or compute cluster may route
through additional component types, depending on their services.

\begin{table}[t!]
\center
\caption{CCNx Name Segment Types}
\begin{tabular}{|p{5cm}|p{9cm}|} \hline
{\bf Name} & {\bf Description} \\ \hline
Name Segment & A generic name segment that includes arbitrary octets. \\
Interest Payload ID & An octet string that identifies the payload carried in an Interest. As an example, the Payload ID might be a hash of the Interest Payload.
This provides a way to differentiate between Interests based on the Payload solely through a Name Segment without having to include all the extra bytes of the payload itself. \\
Application Components &  An application-specific payload in a name segment. An application may apply its own semantics to these components. A good practice is to identify the
application in a Name segment prior to the application component segments. \\ \hline
\end{tabular}
\end{table}

At the lowest level, a Forwarder does not need to understand the
semantics of name segments; it need only identify name segment
boundaries and be able to compare two name segments (both label and
value) for equality.  The Forwarder matches paths segment-by-segment
against its forwarding table to determine a next hop.

\subsection{Name Examples}
This section uses a URI representation of CCNx names. Each component of a name has
a type and value. Examples of this encoding are in Table \ref{tab:names}.

\begin{table}[t!]
\center
\caption{CCNx Name Examples}
\begin{tabular}{|p{5cm}|p{9cm}|} \hline
{\bf Name} & {\bf Description} \\ \hline
ccnx:/ & A 0-length name, corresponds to a default route. \\
ccnx:/NAME= & A name with 1 segment of 0 length, distinct from ccnx:/. \\
ccnx:/NAME=foo/APP:0=bar & A 2-segment name, where the first segment is of type NAME and the second segment is of type APP:0. \\ \hline
\end{tabular}
\label{tab:names}
\end{table}

\subsection{Interest Payload ID}
An Interest may also have a Payload which carries state about the
Interest but is not used to match a Content Object.  If an Interest
contains a payload, the Interest name should contain an Interest
Payload ID (IPID).  The IPID allows a PIT table entry to correctly
multiplex Content Objects in response to a specific Interest with a
specific payload ID.  The IPID could be derived from a hash of the
payload or could be a GUID or a nonce.  An optional Metadata field
defines the IPID field so other systems could verify the IPID, such
as when it is derived from a hash of the payload.  No system is
required to verify the IPID.

\section{Cache Control}
CCNx supports two fields that affect cache control.  These determine
how a cache or Content Store handles a Content Object.  They are not
used in the fast path, but only to determine if a ContentObject can
be injected on to the fast path in response to an Interest.

The ExpiryTime is a field that exists within the signature envelope
of a Validation Algorithm.  It is the UTC time in milliseconds after
which the ContentObject is considered expired and MUST no longer be
used to respond to an Interest from a cache.  Stale content MAY be
flushed from the cache.

The Recommended Cache Time (RCT) is a field that exists outside the
signature envelope.  It is the UTC time in milliseconds after which
the publisher considers the Content Object to be of low value to
cache.  A cache SHOULD discard it after the RCT, though it MAY keep
it and still respond with it.  A cache is MAY discard the content
object before the RCT time too; there is no contractual obligation to
remember anything.

This formulation allows a producer to create a Content Object with a
long ExpiryTime but short RCT and keep re-publishing the same,
signed, Content Object over and over again by extending the RCT.
This allows a form of "phone home" where the publisher wants to
periodically see that the content is being used.

\section{Restrictions}

\subsection{Content Object Hash}
CCNx allows an Interest to restrict a response to a specific hash.
The hash covers the Content Object message body and the validation
sections, if present.  Thus, if a Content Object is signed, its hash
includes that signature value.  The hash does not include the fixed
or hop-by-hop headers of a Content Object.  Because it is part of the
matching rules (see Section 9), the hash is used at every hop.

There are two options for matching the content object hash
restriction in an Interest.  First, a forwarder could compute for
itself the hash value and compare it to the restriction.  This is an
expensive operation.  The second option is for a border device to
compute the hash once and place the value in a header (ConObjHash)
that is carried through the network.  The second option, of course,
removes any security properties from matching the hash, so SHOULD
only be used within a trusted domain.  The header SHOULD be removed
when crossing a trust boundary.

\subsection{Key ID Restriction}
In addition to content restrictions, CCNx allows an Interest to also
restrict a response to a content object
which can be authenticated using a specific public key. This is done by
specifying the identity of the verifying public key in a header (KeyIdRestr)
that is carried through the network. An Interest with a KeyIdRestr only matches a
Content Object if the latter carries a public key whose identity matches the
KeyIdRestr value. An Interest may carry both a content object hash restriction
and a key ID restriction. The former simply subsumes the latter since, by design,
the public key in a matching Content Object would be included in the hash computation input.

\section{Link}
A Link is the tuple
$$
{\mathsf{Name}, [\mathsf{KeyIdRestr}], [\mathsf{ContentObjectHashRestr}]}.
$$
The information in a Link comprises the fields the fields of an
Interest which would retrieve the Link target.  A Content Object with
PayloadType = "Link" is an object whose payload is one or more Links.
This tuple may be used as a KeyLink to identify a specific object
with the certificate wrapped key.  It is RECOMMENDED to include at
least one of KeyIdRestr or ContentObjectHashRestr.  If neither
restriction is present, then any Content Object with a matching name
from any publisher could be returned.

\section{Hashes}
Several protocol fields use cryptographic hash functions, which must
be secure against attack and collisions.  Because these hash
functions change over time, with better ones appearing and old ones
falling victim to attacks, it is important that a CCNx protocol
implementation support hash agility.

In this document, we suggest certain hashes (e.g., SHA-256), but a
specific implementation may use what it deems best.  The normative
CCNx Messages \cite{messages} specification should be taken as the
definition of acceptable hash functions and uses.

\section{Validation}
The Validator consists of a ValidationAlgorithm that specifies how to
verify the message and a ValidationPayload containing the validation
output, e.g., the digital signature or MAC.  The ValidationAlgorithm
section defines the type of algorithm to use and includes any
necessary additional information.  The validation is calculated from
the beginning of the CCNx Message through the end of the
ValidationAlgorithm section.  The ValidationPayload is the integrity
value bytes, such as a MAC or signature.

Some Validators contain a KeyId, identifying the publisher
authenticating the Content Object.  If an Interest carries a
KeyIdRestriction, then that KeyIdRestriction MUST exactly match the
Content Object's KeyId.

Validation Algorithms fall into three categories: MICs, MACs, and
Signatures.  Validators using MIC algorithms do not need to provide
any additional information; they may be computed and verified based
only on the algorithm (e.g., CRC32C).  MAC validators require the use
of a KeyId identifying the secret key used by the authenticator.
Because MACs are usually used between two parties that have already
exchanged secret keys via a key exchange protocol, the KeyId may be
any agreed-upon value to identify which key is used.  Signature
validators use public key cryptographic algorithms such as RSA, DSA,
ECDSA.  The KeyId field in the ValidationAlgorithm identifies the
public key used to verify the signature.  A signature may optionally
include a KeyLocator, as described above, to bundle a Key or
Certificate or KeyLink.  MAC and Signature validators may also
include a SignatureTime, as described above.

A PublicKeyLocator KeyLink points to a Content Object with a DER-
encoded X509 certificate in the payload.  In this case, the target
KeyId must equal the first object's KeyId.  The target KeyLocator
must include the public key corresponding to the KeyId.  That key
must validate the target Signature.  The payload is an X.509
certificate whose public key must match the target KeyLocator's key.
It must be issued by a trusted authority, preferably specifying the
valid namespace of the key in the distinguished name.

\section{Interest to Content Matching} \label{sec:matching}
A Content Object satisfies an Interest if and only if (a) the Content
Object name, if present, exactly matches the Interest name, and (b)
the ValidationAlgorithm KeyId of the Content Object exactly equals
the Interest KeyIdRestriction, if present, and (c) the computed
ContentObjectHash exactly equals the Interest
ContentObjectHashRestriction, if present.

The matching rules are given by this predicate, which if it evaluates
true means the ContentObject matches the Interest.  $N_i$ = Name in
Interest (may not be empty), $K_i$ = KeyIdRestriction in the interest
(may be empty), $H_i$ = ContentObjectHashRestriction in Interest (may be
empty).  Likewise, $N_o$, $K_o$, $H_o$ are those properties in the
ContentObject, where $N_o$ and $K_o$ may be empty; $H_o$ always exists.

As a special case, if the ContentObjectHashRestriction in the
Interest specifies an unsupported hash algorithm, then no
ContentObject can match the Interest so the system should drop the
Interest and MAY send an InterestReturn to the previous hop.  In this
case, the predicate below will never get executed because the
Interest is never forwarded.  If the system is using the optional
behavior of having a different system calculate the hash for it, then
the system may assume all hash functions are supported and leave it
to the other system to accept or reject the Interest.

$$
(\neg N_o \lor (N_i=N_o)) \land (\neg K_i \lor (K_i=K_o)) \land (\neg H_i \lor (H_i=H_o)) \land (\exists N_o \lor \exists H_i)
$$

As one can see, there are two types of attributes one can match.  The
first term depends on the existence of the attribute in the
ContentObject while the next two terms depend on the existence of the
attribute in the Interest.  The last term is the "Nameless Object"
restriction which states that if a Content Object does not have a
Name, then it must match the Interest on at least the Hash
restriction.

If a Content Object does not carry the ContentObjectHash as an
expressed field, it must be calculated in network to match against.
It is sufficient within an autonomous system to calculate a
ContentObjectHash at a border router and carry it via trusted means
within the autonomous system.  If a Content Object
ValidationAlgorithm does not have a KeyId then the Content Object
cannot match an Interest with a KeyIdRestriction.

\section{Interest Return}
This section describes the process whereby a network element may
return an Interest message to a previous hop if there is an error
processing the Interest.  The returned Interest may be further
processed at the previous hop or returned towards the Interest
origin.  When a node returns an Interest it indicates that the
previous hop should not expect a response from that node for the
Interest -- i.e., there is no PIT entry left at the returning node.

The returned message maintains compatibility with the existing TLV
packet format (a fixed header, optional hop-by-hop headers, and the
CCNx message body).  The returned Interest packet is modified in only
two ways:

\begin{itemize}
\item  The PacketType is set to InterestReturn to indicate a Feedback
   message.

\item  The ReturnCode is set to the appropriate value to signal the
   reason for the return
\end{itemize}

The specific encodings of the Interest Return are specified in \cite{messages}.

A Forwarder is not required to send any Interest Return messages.

A Forwarder is not required to process any received Interest Return
message.  If a Forwarder does not process Interest Return messages,
it SHOULD silently drop them.

The Interest Return message does not apply to a Content Object or any
other message type.

An Interest Return message is a 1-hop message between peers.  It is
not propagated multiple hops via the FIB.  An intermediate node that
receives an InterestReturn may take corrective actions or may
propagate its own InterestReturn to previous hops as indicated in the
reverse path of a PIT entry.

\subsection{Message Format}
The Interest Return message looks exactly like the original Interest
message with the exception of the two modifications mentioned above.
The PacketType is set to indicate the message is an InterestReturn
and the reserved byte in the Interest header is used as a Return
Code.  The numeric values for the PacketType and ReturnCodes are in
\cite{messages}.

\subsection{ReturnCode Types}
This section defines the InterestReturn ReturnCode introduced in this
RFC.  The numeric values used in the packet are defined in
\cite{messages}.

\begin{table}[t!]
\center
\begin{tabular}{|p{5cm}|p{9cm}|} \hline
{\bf Name} & {\bf Description} \\ \hline
No Route & The returning Forwarder has no route to the Interest name. \\
HopLimit Exceeded & The HopLimit has decremented to 0 and need to forward the packet. \\
Interest MTU too large & The Interest's MTU does not conform to the require minimum and would require fragmentation. \\
No Resources & The node does not have the resources to process the Interest. \\
Path error & There was a transmission error when forwarding the Interest along a route (a transient error). \\
Prohibited & An administrative setting prohibits processing this Interest. \\
Congestion & The Interest was dropped due to congestion (a transient error). \\
Unsupported Content Object Hash Algorithm & The Interest was dropped because it requested a Content Object Hash Restriction using a hash algorithm that cannot be computed. \\
Malformed Interest & The Interest was dropped beause it did not correctly parse. \\ \hline
\end{tabular}
\end{table}

\subsection{Interest Return Protocol}
This section describes the Forwarder behavior for the various Reason
codes for Interest Return.  A Forwarder is not required to generate
any of the codes, but if it does, it MUST conform to this
specification.

If a Forwarder receives an Interest Return, it SHOULD take these
standard corrective actions.  A forwarder is allowed to ignore
Interest Return messages, in which case its PIT entry would go
through normal timeout processes.

\begin{itemize}
\item  Verify that the Interest Return came from a next-hop to which it
   actually sent the Interest.

\item  If a PIT entry for the corresponding Interest does not exist, the
   Forwarder should ignore the Interest Return.

\item  If a PIT entry for the corresponding Interest does exist, the
   Forwarder MAY do one of the following:
\begin{itemize}
   \item  Try a different forwarding path, if one exists, and discard the
      Interest Return, or

   \item Clear the PIT state and send an Interest Return along the
      reverse path.
\end{itemize}
\end{itemize}

If a forwarder tries alternate routes, it MUST ensure that it does
not use same same path multiple times.  For example, it could keep
track of which next hops it has tried and not re-use them.

If a forwarder tries an alternate route, it may receive a second
InterestReturn, possibly of a different type than the first
InterestReturn.  For example, node A sends an Interest to node B,
which sends a No Route return.  Node A then tries node C, which sends
a Prohibited.  Node A should choose what it thinks is the appropriate
code to send back to its previous hop

If a forwarder tries an alternate route, it should decrement the
Interest Lifetime to account for the time spent thus far processing
the Interest.

\subsubsection{No Route}
If a Forwarder receives an Interest for which it has no route, or for
which the only route is back towards the system that sent the
Interest, the Forwarder SHOULD generate a "No Route" Interest Return
message.

How a forwarder manages the FIB table when it receives a No Route
message is implementation dependent.  In general, receiving a No
Route Interest Return should not cause a forwarder to remove a route.
The dynamic routing protocol that installed the route should correct
the route or the administrator who created a static route should
correct the configuration.  A forwarder could suppress using that
next hop for some period of time.

\subsubsection{HopLimit Exceeded}
A Forwarder MAY choose to send HopLimit Exceeded messages when it
receives an Interest that must be forwarded off system and the
HopLimit is 0.

\subsubsection{Interest MTU Too Large}
If a Forwarder receives an Interest whose MTU exceeds the prescribed
minimum, it MAY send an "Interest MTU Too Large" message, or it may
silently discard the Interest.

If a Forwarder receives an "Interest MTU Too Large" is SHOULD NOT try
alternate paths. It SHOULD propagate the Interest Return to its
previous hops.

\subsubsection{No Resources}
If a Forwarder receives an Interest and it cannot process the
Interest due to lack of resources, it MAY send an InterestReturn. A
lack of resources could be the PIT table is too large, or some other
capacity limit.

\subsubsection{Path Error}
If a forwarder detects an error forwarding an Interest, such as over
a reliable link, it MAY send a Path Error Interest Return indicating
that it was not able to send or repair a forwarding error.

\subsubsection{Prohibited}
A forwarder may have administrative policies, such as access control
lists, that prohibit receiving or forwarding an Interest.  If a
forwarder discards an Interest due to a policy, it MAY send a
Prohibited InterestReturn to the previous hop.  For example, if there
is an ACL that says /parc/private can only come from interface e0,
but the Forwarder receives one from e1, the Forwarder must have a way
to return the Interest with an explanation.

\subsubsection{Congestion}
If a forwarder discards an Interest due to congestion, it MAY send a
Congestion InterestReturn to the previous hop.

\subsubsection{Unsupported Content Object Hash Algorithm}
If a Content Object Hash Restriction specifies a hash algorithm the
forwarder cannot verify, the Interest should not be accepted and the
forwarder MAY send an InterestReturn to the previous hop.

\subsubsection{Malformed Interest}
If a forwarder detects a structural or syntactical error in an
Interest, it SHOULD drop the interest and MAY send an InterestReturn
to the previous hop.  This does not imply that any router must
validate the entire structure of an Interest.

\bibliographystyle{unsrt}
\bibliography{references}

\end{document}